\documentclass[]{aa}
\usepackage{graphicx}
\usepackage{txfonts}
\begin{document}
%%%%%%%%%%%%%%%%%%%%%%%%%%%%%%%%%%%%%%%%%%%%%%%%%%%%%%%%%%%%%%%%%%%%%%

   \title{The influence of the Hall effect on the global stability of cool
protostellar disks}

   \titlerunning{Hall effect and the global stability of protostellar disks }
\authorrunning{G. R\"udiger \& L. L. Kitchatinov}
   \author{G. R\"udiger\inst{1,3} \and L. L. Kitchatinov\inst{1,2}}

  % \offprints{gruediger@aip.de}

   \institute{Astrophysikalisches Institut Potsdam, An der Sternwarte 16,
              D-14482, Potsdam, Germany \\
              \email{gruediger@aip.de}
         \and
             Institute for Solar-Terrestrial Physics, PO Box
             4026, Irkutsk 664033, Russian Federation\\
             \email{kit@iszf.irk.ru}
              \and Isaac Newton Institute for Mathematical Sciences, 20 Clarkson Rd, Cambridge, CB3 0EH, U.K.}

   \date{\today}

%%%%%%%%%%%%%%%%%%%%%%%%%%%%%%%%%%%%%%%%%%%%%%%%%%%%%%%%%%%%%%%%%%%%%%%
\abstract{
The influence of the Hall effect on the global stability of cool Kepler disks under the influence of an axial magnetic field is considered. For sufficiently large magnetic Reynolds numbers Rm the magnetorotational instability (MRI) exists in a finite interval  of magnetic field amplitudes, $B_{\rm min} < B < B_{\rm max}$. For Kepler  disks the pure MRI needs both rather high Rm (representing the needed electrical conductivity) as well as   $B_{\rm min}$ of order  0.1 G.  The magnetic field pattern resulting from  our global and linear calculations is of quadrupolar parity. For magnetic  fields {\em antiparallel} to the rotation axis the Hall effect  { reduces} the minimum magnetic Reynolds number by about one order of magnitude.  The   $B_{\min}$, however, is even (sightly) increased (see Fig. \ref{f5}).\\
For magnetic  fields {\em parallel} to the rotation axis the Hall
effect drives its own instability without the action of the Lorentz force. The
corresponding critical  magnetic Reynolds number  proves to be  larger with the
Hall effect (${\rm Rm}\sim 10$) than without the Hall effect (${\rm Rm}\sim 7$) so that  the Hall effect  for parallel fields even  disturbs the  formation of MHD-instability  in cool protoplanetary disks. If the disk is supercritical then the main result of the Hall effect for positive  fields  is the strong reduction of the minimum magnetic field amplitude which is necessary to start the instability. Observations must show whether in star-forming regions the rotation axis and the magnetic field orientation are correlated or are anticorrelated.
If the magnetic fields are large enough then our model predicts the dominance of fields antiparallel to the rotation axis.
%%%%%%%%%%%%%%%%%%%%%%%%%%%%%%%%%%%%%%%%%%%%%%%%%%%%%%%%%%%%%%%%%%%%%%%
}
%%%%%%%%%%%%%%%%%%%%%%%%%%%%%%%%%%%%%%%%%%%%%%%%%%%%%%%%%%%%%%%%%%%%%%%
\maketitle
%%%%%%%%%%%%%%%%%%%%%%%%%%%%%%%%%%%%%%%%%%%%%%%%%%%%%%%%%%%%%%%%%%%%%%%
\section{Introduction}
%%%%%%%%%%%%%%%%%%%%%%%%%%%%%%%%%%%%%%%%%%%%%%%%%%%%%%%%%%%%%%%%%%%%%%%
The Hall effect in protostellar disks, with their low degree of ionization, has recently become
a subject of increasing interest due to its relevance to the stability
and the angular momentum transport in the disks. The Hall effect can amplify
or suppress the standard magnetorotational instability
(hereafter MRI, see Balbus \& Hawley 1991)  depending on the sign of the product of angular velocity and magnetic field projections on the wave vector of a disturbance.
The effect was found
to destabilize when the product is \emph{negative} (Wardle 1999; Balbus \& Terquem 2001; R\"udiger \& Shalybkov 2004). In this paper we shall present a study for the global stability of a differentially rotating disk of given (small) thickness and (low) temperature.

As the solution of the induction equation alone, the Hall effect can drive its 
own instability. The instability does not require rotation; it also exists for 
a plane shear flow. For 
${\rm d}\Omega/{\rm d}R<0$ this shear-Hall instability develops when the axial 
magnetic field is \emph{positive} and vice versa. Similar to MRI,  it only 
exists between a minimum field,  $B_{\rm min}$, and a maximum field,  $B_{\rm max}$. 
For a positive magnetic field, MRI and  the Hall effect amplify each other close to $B_{\rm min}$, and they compete close to $B_{\rm max}$.  Both boundaries of the instability range are thus reduced  by the interplay of the two effects in this case.

For negative axial fields  the Hall effect   and MRI amplify each other when 
the field strength is close to $B_{\rm max}$. The Hall effect is  
destabilizing here as  it transforms  the value of $B_{\rm max}$ to higher 
values. The Hall effect and MRI compete, however,  close to $B_{\rm min}$. Here, the  Hall effect is stabilizing as the  $B_{\rm min}$ is increased. Both boundaries of the instability range are thus  enhanced  by the interplay of the two effects in this case.

%%%%%%%%%%%%%%%%%%%%%%%%%%%%%%%%%%%%%%%%%%%%%%%%%%%%%%%%%%%%%%%%%%%%%%%
\section{Local approximation}\label{guidence}
%%%%%%%%%%%%%%%%%%%%%%%%%%%%%%%%%%%%%%%%%%%%%%%%%%%%%%%%%%%%%%%%%%%%%%%
\subsection{Linearized equations}
%%%%%%%%%%%%%%%%%%%%%%%%%%%%%%%%%%%%%%%%%%%%%%%%%%%%%%%%%%%%%%%%%%%%%%%
The conditions in protostellar disks were discussed by
Balbus \& Terquem (2001). The disk material is partly ionized plasma where ions are well linked to neutrals but
electrons are not. This leads to the induction equation including an
additional term compared to the standard one-fluid MHD, i.e.
\begin{equation}
 {\partial{\vec B}\over\partial t}\ = \ {\nabla\times} \left(
 {\vec u}\times{\vec B}\ +\ {\vec u}_{\rm H}\times{\vec B}\
 -\ \eta\ {\nabla\times}{\vec B}
 \right),
\label{1.1}
\end{equation}
where the second term on the right stands for the Hall electromotive force with the
effective velocity, ${\vec u}_{\rm H}$, proportional to the current density, ${\vec J}={\nabla\times}{\vec B}/\mu_0 $:
\begin{eqnarray}
 {\vec u}_{\rm H} &=& -{{\vec J}\over e n_{\rm e}}\ =
 - \eta\ C_{\rm H} {\nabla\times{\vec B}\over B} ,
 \label{1.2} \\
 C_{\rm H} &=& {\omega_{\rm ce}\over \nu_{\rm e}} .
 \label{1.3}
\end{eqnarray}
In these equations, $n_{\rm e}$ is the electron number density, 
$\omega_{\rm ce} = eB/c m_{\rm e}$ is the cyclotron frequency, and $\nu_{\rm e}$ is collision frequency of electrons. The reason to introduce the \lq Hall velocity' (\ref{1.2}) is that it helps in interpreting future results. Depending on the magnetic field direction the quantity $C_{\rm H}$ can be positive or negative.

The equation of motion reads
\begin{equation}
  {\partial{\vec u}\over\partial t}\
  +\ \left({\vec u}\cdot\nabla\right){\vec u}\ =\ -{1\over\rho}\nabla P\
  -\ \nabla\Phi\ +\ {1\over\rho}{\vec J}\times{\vec B}\ +\ \nu\Delta{\vec u},
\label{1.4}
\end{equation}
where the viscosity term is kept for numerical reasons although the viscosity is
small for protostellar disks. We shall see that the stability parameters do not depend on the viscosity whenever the magnetic Prandtl number,
\begin{equation}
 {\rm Pm}\ =\ {\nu\over\eta} ,
\label{1.5}
\end{equation}
is below $0.1$.
We assume an incompressible fluid, ${\rm div}\,{\vec u} = 0$. 

The reference state includes a non-uniform rotation with the
angular velocity, $\Omega$, dependent on the distance $s$ to the rotation axis, and a uniform axial magnetic field ${\vec B}_0\|{\vec\Omega}$. Linear stability or instability against small disturbances is considered.

For a local approximation 
 a Cartesian coordinate system rotating with the local angular
velocity $\Omega$ is used with  $x,y$ and $z$ pointing in the radial, 
azimuthal and vertical directions (see Balbus \& Hawley 1991; Brandenburg et al. 1995). The local approximation concerns perturbations whose spatial scales are small compared to the global scale  of the disk parameters. The  rotation law can then be approximated by the shear flow
${\vec U}_0 = -\hat{\vec e}_y\Omega q x$
where $q$ is the (constant) local shear. The linearized MHD equations
with the Hall effect  read
\begin{eqnarray}
\lefteqn{   {\partial{\vec B'}\over\partial t} -B_0{\partial
   {\vec u'}\over\partial z} -
   x q \Omega{\partial{\vec B'}\over\partial y}
   + q\Omega B'_x \hat{\vec e}_y -}
   \nonumber \\
   && \quad \quad \quad \quad - \eta\Delta{\vec B'} +
   \eta C_{\rm H}{\partial\left({\nabla\times}{\vec B'}\right)
   \over \partial z} = 0,
   \nonumber \\
\lefteqn{   {\partial{\vec u'}\over\partial t} + 2\Omega\hat{\vec e}_z\times{\vec u'}
   - x q \Omega{\partial{\vec u'}\over\partial y} - q\Omega u'_x \hat{\vec e}_y
   -}
   \nonumber \\
   && \quad \quad \quad \quad - {B_0\over 4\pi \rho}{\partial{\vec B'}\over\partial z} + {1\over\rho}
   \nabla P' - \nu\Delta{\vec u'} = 0 
\label{1.6}
\end{eqnarray}
with dashes indicating the small disturbances.

Considering plane waves with ${\vec B}',{\vec u}',P'\sim {\rm
exp}(\gamma t + {\rm i}k z)$ leads to the dispersion relation
\begin{eqnarray}
  \lefteqn{\left(\left(\gamma + \eta k^2\right)^2 + \omega_{\rm H}
  \left(\omega_{\rm H} - q\Omega\right)\right)
  \left(\left(\gamma + \nu k^2\right)^2 + 2\left( 2 - q\right)\Omega^2\right)+}
  \nonumber \\
  && \quad \quad \quad +\ \omega_{\rm A}^2\left( \omega_{\rm A}^2 - 2 q \Omega^2
   + 2\left(\gamma + \nu k^2\right)\left(\gamma +\eta k^2\right)
   \right)+
  \nonumber \\
  && \quad \quad \quad + \left( 4 - q\right) \omega_{\rm A}^2\omega_{\rm H}\ \Omega\
  =\ 0 ,
\label{1.7}
\end{eqnarray}
with the Alfv\'en and the Hall
frequencies
\begin{equation}
  \omega_{\rm A} =\ kB_0/\sqrt{\mu_0 \rho}, \ \ \ \
  \omega_{\rm H} =\ \eta k^2 C_{\rm H} .
  \label{1.8}
\end{equation}
If overstable modes are ignored, $\gamma = 0$ in (\ref{1.7}) gives the equation 
\begin{eqnarray}
 \lefteqn{ \left( 1 + C_{\rm H}\left( C_{\rm H} - q
 {\rm Rm}\right)\right)
  \left( 1 + 2\left( 2- q \right){{\rm Rm}^2\over{\rm Pm}^2}\right)+}
  \nonumber \\
  && \quad +\ {{\rm Ha}^2{\rm Rm}\over{\rm Pm}}\left(\left( 4- q\right)
  C_{\rm H}
  - 2 q {\rm Rm}\right) + 2 {\rm Ha}^2 + {\rm Ha}^4 = 0.
  \label{1.9}
\end{eqnarray}
for the marginal stability
 separating the regions of instability ($\Re(\gamma) > 0$) and stability
($\Re (\gamma ) < 0$). Here  Rm and Ha are the local magnetic Reynolds number and the Hartmann
number
\begin{equation}
  {\rm Rm} = {\Omega\over\eta k^2},\ \ \ \ \ \ \ \ \ \ \ \ \ \ \ \ \  {\rm Ha} =      {\omega_{\rm A}\over k^2\sqrt{\eta\nu}},
\label{1.10}
\end{equation}
both taken as positive-definite throughout the paper. The shear is obviously necessary for any instability
because  Eq. (\ref{1.9})  provides solutions with
real $C_{\rm H}$ and Ha only with finite values of $q$ and Rm.
%%%%%%%%%%%%%%%%%%%%%%%%%%%%%%%%%%%%%%%%%%%%%%%%%%%%%%%%%%%%%%%%%%%%%%%
\subsection{Shear-Hall instability}\label{SH}
%%%%%%%%%%%%%%%%%%%%%%%%%%%%%%%%%%%%%%%%%%%%%%%%%%%%%%%%%%%%%%%%%%%%%%%
Consider the case where the
Hall frequency (\ref{1.8}) is large,
\begin{equation}
 \omega_{\rm H}^2 \gg \omega_{\rm A}^2.
 \label{1.11}
\end{equation}
Then the dispersion relation (\ref{1.7})
 reduces to
\begin{equation}
 \left(\gamma + \eta k^2\right)^2 + \omega_{\rm H}
  \left(\omega_{\rm H} - q\Omega\right)\ =\ 0.
\label{1.12}
\end{equation}
The induction equation has been 
decoupled from the equation of motion, and the dispersion relation (\ref{1.12}) can be found from the induction equation alone. An  instability exists if  the magnetic Reynolds number exceeds the minimum value $2/q$.
%\begin{equation}
 % {\rm Rm}_{\rm min} = 2/q .
 % \label{1.13}
%\end{equation}
If $q$ is positive it  exists for positive $C_{\rm H}$, i.e. for the external
magnetic field parallel to the angular velocity, and vice versa ($q\omega_{\rm H}$ must be positive, see (\ref{1.12})). For large ${\rm Rm} \gg {\rm Rm}_{\rm min}$, the instability region is given by 
\begin{equation}
  {1\over q{\rm Rm}} < C_{\rm H} < q {\rm Rm}. 
  \label{1.14}
\end{equation}
 The Hall parameter (\ref{1.3}) is
proportional to the magnetic field. Similar to MRI, the shear-Hall instability
exists in a limited range of magnetic fields. In a further similarity, the maximum growth rate is controlled by the local Oort A value, i.e.
\begin{equation}
  \gamma_{\rm max} + \eta k^2 =\ q\Omega /2 .
\label{1.15}
\end{equation}

%%%%%%%%%%%%%%%%%%%%%%%%%%%%%%%%%%%%%%%%%%%%%%%%%%%%%%%%%%%%%%%%%%%%%%%
\subsection{MRI for low conductivity, ${\rm Pm} \ll 1$}
%%%%%%%%%%%%%%%%%%%%%%%%%%%%%%%%%%%%%%%%%%%%%%%%%%%%%%%%%%%%%%%%%%%%%%%
For  the  small magnetic Prandtl numbers expected for protostellar disks an
appropriate scaling  of the MRI parameters exists. Without the Hall effect  ($C_{\rm H} = 0$) Eq.  (\ref{1.9}) yields
the neutral stability condition
\begin{equation}
   {\rm Rm}^2 = {{\rm Pm}^2\left( 1 + {\rm Ha}^2\right)^2
   \over 2\left( q \ {\rm Pm}\ {\rm Ha}^2 - 2+q \right)} 
\label{1.16}
\end{equation}
 for MRI alone. Suppose that Pm is decreased by decreasing the viscosity but 
 keeping the  magnetic diffusivity finite. Then the Lundquist number
\begin{equation}
  {\rm S} = \sqrt{\rm Pm}\ {\rm Ha} = {\omega_{\rm A}\over\eta k^2} ,
  \label{1.17}
\end{equation}
remains finite. At small Pm, Eq. (\ref{1.16}) becomes
\begin{equation}
  {\rm Rm}^2 = {{\rm S}^4
   \over 2\left( q \ {\rm S}^2 - 2+q \right)} .
   \label{1.18}
\end{equation}
This equation does not include Pm which means that the actual viscosity value is
not important provided that it is smaller than the magnetic diffusivity.
We shall see in Sect.~\ref{results} that  a representation of MRI in terms of both the magnetic Reynolds and Lundquist numbers becomes independent of Pm when
the latter drops below (say) $0.1$. In a global model, therefore, a small but finite viscosity can be kept for numerical stability to produce the results which remain valid for arbitrarily small Pm when represented in terms of  Rm and S.
Hereafter, we always use the Lundquist number (\ref{1.17}) rather than  Ha.
%%%%%%%%%%%%%%%%%%%%%%%%%%%%%%%%%%%%%%%%%%%%%%%%%%%%%%%%%%%%%%%%%%%%%%%
\subsection{MRI plus Hall effect  for $\rm Pm=1$}
%%%%%%%%%%%%%%%%%%%%%%%%%%%%%%%%%%%%%%%%%%%%%%%%%%%%%%%%%%%%%%%%%%%%%%%
Consider the  interplay of MRI and the Hall effect for the  simplifying case of Pm=1.  Then the  marginal stability equation (\ref{1.9}) becomes
\begin{eqnarray}
  {\rm S}^2 +\ 2 C_{\rm H}{\rm Rm}\ +\ 1
  &=& {q\over 2}{\rm Rm}\left( 2{\rm Rm} + C_{\rm H}\right)
  \nonumber \\
  &\mp& \left( 2{\rm Rm} - C_{\rm H}\right)
  \left( {q^2{\rm Rm}^2\over 4} - 1\right)^{1/2} .
  \label{1.19}
\end{eqnarray}
The plus and minus signs on  the RHS of this equation define two boundaries of
the instability region for a given Rm  provided that  the magnetic Reynolds number  exceeds the minimum value of ${\rm Rm}_{\rm min} = 2/q$.

Note that the Hall parameter (\ref{1.3}) and the Lundquist number (\ref{1.17}) are both linear in the  external  field $B_0$. Their ratio
\begin{equation}
  \beta =  \frac{C_{\rm H}} {\rm S},
  \label{1.20}
\end{equation}
therefore, characterizes mainly the material. It may be  taken to be positive 
for magnetic fields parallel to the rotation axis and negative  for magnetic fields antiparallel to the rotation axis. 

For both cases the results are quite different.  Consider the instability region for very large
${\rm Rm} \gg {\rm max}\left({\rm Rm}_{\rm min},\beta^{-1}\right)$. For {\em positive} $\beta$  it is
\begin{equation}
   {1\over\beta q{\rm Rm}} < {\rm S} <
   {\rm Rm}\left(\left( 2 q + \beta^2\right)^{1/2} - \beta\right).
   \label{1.21}
\end{equation}
The lower limit is the same as in  (\ref{1.14}). $B_{\rm min}$ for positive $\beta$ is thus controlled by the shear-Hall instability.
The lower  bound is small compared to ${\rm S}_{\rm min} = \sqrt{(2-q)/q}$ expected from
Eq.  (\ref{1.18}) for MRI. The Hall effect for positive $B_0$ amplifies MRI
for the fields close to $B_{\rm min}$. The effect is, however, stabilizing 
close to $B_{\rm max}$ because the upper limit in (\ref{1.21})  decreases with $\beta$.
\begin{figure}
   \centering
   \includegraphics[width=7cm]{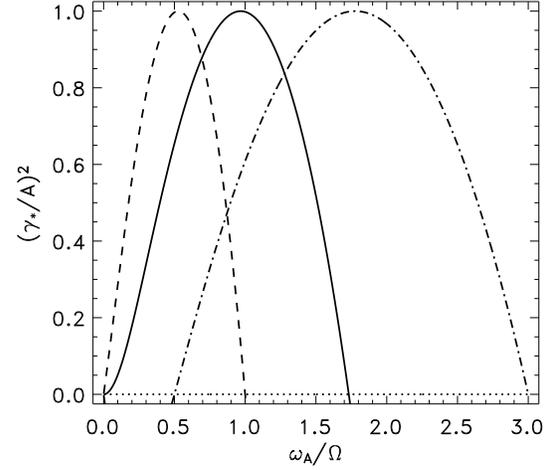}
      \caption{The growth rates, $\gamma_* = \gamma+\eta k^2$ in units
      of A=$q\Omega/2$ as functions of the normalized
      field strength for MRI alone (full line) and MRI modified by the
      Hall effect of positive fields ($\beta = 1$, dashed line) and negative fields 
      ($\beta= -1$, dashed-dotted). ${\rm Pm} =1$.}
   \label{f1}
\end{figure}

The instability domain for {\em negative} $\beta$  is
\begin{equation}
  |\beta|\left( 2 - q\right) {\rm Rm} < {\rm S} <
  \left(\sqrt{2 q + \beta^2} + |\beta|\right) {\rm Rm} .
  \label{1.22}
\end{equation}
Both bounds  increase with $|\beta|$. Here the Hall effect is thus destabilizing close to $B_{\rm max}$ (Wardle 1999; Balbus \& Terquem 2001) and stabilizing close to $B_{\rm min}$.

In  Fig.~\ref{f1} the interaction of MRI with the Hall effect is   demonstrated  by
 the growth rates derived from the dispersion relation (\ref{1.7}) for  ${\rm Pm} = 1$.
The solid line gives the typical growth rate profile for the MRI alone. For positive   fields the profile is moved by the Hall effect to the left and for negative fields the profile is moved to the right. At the strong-field limits only the negative  Hall effect  is thus  destabilizing. In opposition  to that the positive Hall effect even disturbes the instability and may not help to produce the desired turbulence in cool disks.

%%%%%%%%%%%%%%%%%%%%%%%%%%%%%%%%%%%%%%%%%%%%%%%%%%%%%%%%%%%%%%%%%%%%%%%
\section{A global model}\label{model}
%%%%%%%%%%%%%%%%%%%%%%%%%%%%%%%%%%%%%%%%%%%%%%%%%%%%%%%%%%%%%%%%%%%%%%%
The above  results for growth or decay of small disturbances taken from the 
local dispersion relation have been known since the papers of  Wardle (1999) 
and Balbus \& Terquem (2001).   Contrary to the local considerations we shall 
present in the following the results for global Kepler flow models with Ohmic 
dissipation and the Hall effect included. The  progress of such an approach is 
to find exact values for the critical  Reynolds numbers,   the minimal and 
maximal magnetic amplitudes and the global geometry of the resulting 
instabilty pattern. As the limiting magnetic fields prove to be  surprisingly 
high (and rather sensitive to the Hall effect) information about the minimum 
magnetic fields is important for the discussion of the  
(magneto-)hydrodynamics of cool protostellar disks.

Our  global model differs from that of Kitchatinov \& Mazur (1997) only by the inclusion of the Hall effect. 
The model concerns a rotating disk of constant thickness, $2H$,
threaded by a uniform axial magnetic field. The rotation axis is
normal to the disk, and the angular velocity, $\Omega$, depends on
the distance, $R$, to the axis. This dependence is parameterized
by
\begin{equation}
   \Omega (R) = \Omega_0 \tilde\Omega (R)
\label{2.1}
\end{equation}
with
\begin{equation}
   \tilde\Omega (R) = \left( 1 + \left( {R\over R_0}\right)^{3n/2}\
   \right)^{-1/n}.
\label{2.2}
\end{equation}
This profile describes almost uniform rotation at  small distances $R\ll R_0$,
which smoothly transforms to the Keplerian law, $\Omega \simeq \Omega_0 (R_0/R)^{3/2}$, for large distances
$R \gg R_0$. We use $n=2$ in Eq.~(\ref{2.2}) and $R_0/H = 5$ for the aspect ratio and  ${\rm div}\ {\vec u} = 0$ as above.  The pressure is excluded by {\em curling} Eq.   (\ref{1.4}). This yields  for the vorticity, ${\vec\omega} = \nabla\times{\vec u}$, the relation
\begin{equation}
  {\partial{\vec\omega}\over\partial t} =\ {\vec\nabla}\times\left(
  {\vec u}\times{\vec\omega} + {\vec J}\times{\vec B}/\rho\right)
  + \nu\Delta{\vec\omega}.
\label{2.3}
\end{equation}
The Eqs.~(\ref{1.1})
and (\ref{2.3}) are about the rotation (\ref{2.1}) and the
uniform axial field, ${\vec B}_0 = B_0 \hat{\vec{e}}_z$ ($\hat{\vec{e}}_z$ is
the unit vector along the rotation axis), and  the
normalized variables
\begin{equation}
\hat{\vec b} = {\vec B'}/B_0,
\ \hat{\vec j} = {\mu_0  H\over  B_0}{\vec J'},
\ \hat{\vec u} = {\vec u'}/(H\Omega_0),
\ \hat{\vec\omega} = {\vec\omega '}/\Omega_0 ,
\label{2.4}
\end{equation}
for the disturbances are introduced. 
This leads to four basic dimensionless parameters among which the Hall parameter, $C_{\rm H}$,  and magnetic Prandtl number, Pm,   are defined by  (\ref{1.3}) and (\ref{1.5}) but magnetic Reynolds number and the Lundquist number now are  
\begin{equation}
   {\rm Rm} = {\Omega_0 H^2\over\eta},
   \ \ \ \ \ \ \
   {\rm S} = {B_0 H\over \sqrt{\mu_0 \rho}\eta}.
   \label{2.5}
\end{equation}
These parameters control the equation system for the normalized disturbances
\begin{eqnarray}
   {\partial\hat{\vec b}\over\partial t} &=& {\rm Rm}\ \nabla\times\left(
   R \tilde \Omega(R)\ \hat{\vec e}_\phi\times\hat{\vec b} -
    \hat{\vec e}_z\times\hat{\vec u}
   \right) - C_{\rm H}\left(\hat{\vec e}_z\cdot\nabla\right)\hat{\vec j}
   + \Delta\hat{\vec b} ,
   \nonumber \\
   {\partial\hat{\vec\omega}\over\partial t} &=&
   {\rm Rm}\ \nabla\times\left(
   R \tilde\Omega(R)\ \hat{\vec e}_\phi\times\hat{\vec\omega} -
   {\kappa^2\over 2\tilde\Omega (R)}\
   \hat{\vec e}_z\times\hat{\vec u}\right)
   \nonumber \\
   &+& {{\rm S}^2\over{\rm Rm}}\left(
   \hat{\vec e}_z\cdot\nabla\right)\hat{\vec j}
   + {\rm Pm}\Delta\hat{\vec\omega},
\label{2.6}
\end{eqnarray}
where $\kappa$ is the
normalized epicycle frequency
\begin{equation}
  \kappa^2 = {2\tilde\Omega\over R}{{\rm d}
  \left(R^2 \tilde\Omega\right)\over{\rm d} R},
\label{2.7}
\end{equation}
and time and distances are normalized to  the diffusion time, $H^2/\eta$, and
 the disk half-thickness, $H$.  Here only the stability of the rotation law is considered, the accretion  flow which is connected to this nonuniform rotation via the viscosity is neglected 
 (see Kersal\'e et al. 2004).

The boundary conditions on the disk surfaces are (i) stress-free for the
flow and (ii) pseudo-vacuum conditions for the magnetic field fluctuations, i.e.
\begin{equation}
\hat{\vec{e}}_z\times\hat{\vec b} = 0.
\label{2.8}
\end{equation}
The solutions  are required to be regular on the rotation axis and
to vanish at infinity.

The linear stability analysis with $\hat{\vec b}, \hat{\vec\omega} \sim
{\exp}(\gamma t)$ leads to an eigenvalue problem for the equation
set (\ref{2.6}) which has been solved numerically. A new variable, $y$, 
\begin{equation}
  y = {R/R_0\over 1 + R/R_0} , \hspace{1 cm} 0\leq y\leq 1 ,
  \label{2.9}
\end{equation}
has been introduced transforming the infinite disk to a finite domain. 
 $R_0$ is the turnover  radius in Eq.~(\ref{2.2}). A  uniform grid in $y$ was
 applied which corresponds to a non-uniform grid in $R$. The eigenvalues and eigenvectors are computed by the inverse iteration method.

 The system (\ref{2.6}) allows two types of solutions with
different symmetries about the disk midplane. One of the symmetry
types combines a symmetric magnetic field with an antisymmetric
flow field. The notation, Sm, will be used for this type of eigenmodes, where \lq m' is the azimuthal wave number, i.e. S0 represents  an axisymmetric mode, S1 defines the nonaxisymmetric mode with $m =1$, and so on.
The other symmetry type combines
antisymmetric magnetic field with symmetric flow. The notation Am is used for the eigenmodes of this type of symmetry.

The primary goal of the linear theory is to define the stability
boundary in parameter space  which separates the
region of stable perturbations with negative or zero real part of $\gamma$
from the instability region with (exponentially) growing
perturbations. The stability map strongly  depends on the symmetry type of
the excitation.
%%%%%%%%%%%%%%%%%%%%%%%%%%%%%%%%%%%%%%%%%%%%%%%%%%%%%%%%%%%%%%%%%%%%%%%
\section{Results}\label{results}
%%%%%%%%%%%%%%%%%%%%%%%%%%%%%%%%%%%%%%%%%%%%%%%%%%%%%%%%%%%%%%%%%%%%%%%
\subsection{MRI for low conductivity (small Pm) }
\begin{figure}
   \centering
   \includegraphics[height=6cm,width=7cm]{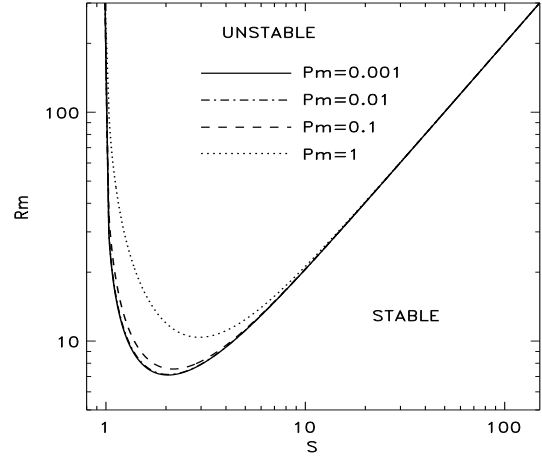}
   \caption{Neutral stability lines of the MRI for various magnetic
            Prandtl numbers ($C_{\rm H}=0$).  For ${\rm Pm} < 0.1$   the lines are almost independent of Pm.}
   \label{f4}
\end{figure}

%%%%%%%%%%%%%%%%%%%%%%%%%%%%%%%%%%%%%%%%%%%%%%%%%%%%%%%%%%%%%%%%%%%%%%%
Consider first with $C_{\rm H}=0$  the  MRI alone.
Figure~\ref{f4} shows the neutral  stability lines for  Prandtl numbers decreasing from Pm=1 to smaller values. The stability  is almost independent of the magnetic Prandtl number for sufficiently small Pm,  say ${\rm Pm} < 0.1$. The MRI characteristics computed with moderately small Pm remain valid for arbitrary small Pm. This finding may be important for predicting MRI parameters for laboratory experiments and protostellar disks where magnetic Prandtl numbers are very small. Hereafter, we fix Pm = 0.01.

The minimum value for Rm for instability in the small Pm regime is
\begin{equation}
   {\rm Rm}_{\rm min} \simeq 7.1.
 \label{3.4}
\end{equation}
For Reynolds numbers large compared to this value the instability exists within the magnetic range
\begin{equation}
  1 < {\rm S} < 0.5\cdot {\rm Rm}.
\label{3.5}
\end{equation}
Cool protostellar disks  may not  reach such  values of the magnetic Reynolds number. On the other hand, for protostellar disks  the minimum condition for the magnetic field (${\rm S}\simeq 1$) proves to be  very stringent (in great contrast to the MRI in galaxies,  see 
Kitchatinov \& R\"udiger 2004). With the values given below for $\eta$ (10$^{15}$ cm$^2$/s),  density (10$^{-10}$ g/cm$^3$) and for a disk height of 0.1 AU a  minimum field of almost  0.1 G  is needed to fulfill ${\rm S}=1$.
It is thus tempting for  different reasons  to probe the shear-Hall instability for protostellar disks which, however,   only exists  for magnetic fields {\em parallel} to the rotation axis.
%%%%%%%%%%%%%%%%%%%%%%%%%%%%%%%%%%%%%%%%%%%%%%%%%%%%%%%%%%%%%%%%%%%%%%%
\subsection{Shear-Hall instability}
%%%%%%%%%%%%%%%%%%%%%%%%%%%%%%%%%%%%%%%%%%%%%%%%%%%%%%%%%%%%%%%%%%%%%%%
Another extreme  exists for magnetic fields parallel to the
rotation axis. The Hall effect for one sign of the magnetic field in connection
with differential rotation can form its own
instability as shown by the solution of the induction equation alone. In the present  model this shear-Hall instability can be found for  small
Lundquist number, S~$\ll C_{\rm H}$, i.e. large $\beta$ of Eq.~(\ref{1.20}). In this case the Lorentz force in (\ref{1.4}) can be neglected so that MRI is excluded.
\begin{figure}
   \centering
   \includegraphics[height=6cm,width=7cm]{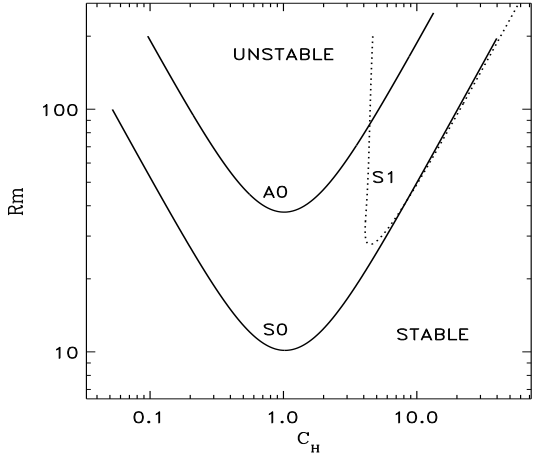}
      \caption{Stability diagram for the shear-Hall instability
      (positive magnetic field, i.e. {\em parallel} to the
      rotation axis; only the induction equation is solved).
      Lines are labeled by their symmetry types.
      The axisymmetric quadrupolar  mode (S0) is preferred.}
   \label{f2}
\end{figure}

Figure~\ref{f2} shows for $\beta \to \infty$ the stability map for modes of 
three basic symmetry types, ie. S0, S1, and A0.
The quadrupolar axisymmetric modes S0 are preferentially excited. The neutral stability lines for the other symmetry types 
lie completely  inside the unstable region of S0.  It is one of the advantages 
of our approach that the equatorial symmetry results from the computations 
rather than needing to be prescribed. Salmeron \& Wardle (2003) in their extensive analysis of the vertical structure of accretion disks  used those symmetry conditions at the equator valid only for A0-fields. When the system becomes unstable  against  A0-disturbances it is thus already unstable against S0-perturbations.

The minimum Reynolds number for all the modes  is
\begin{equation}
 {\rm Rm}_{\rm min} \simeq 10.
\label{3.1}
\end{equation}
For large Rm, the left branch of the neutral stability line for S0-modes
approaches the relation
\begin{equation}
 C_{\rm H} \simeq 5/{\rm Rm},
 \label{3.2}
\end{equation}
while the right branch is close to
\begin{equation}
 C_{\rm H} \simeq  0.2\cdot{\rm Rm}.
 \label{3.3}
\end{equation}
Note that for large magnetic Reynolds number Rm the minimum  $C_{\rm H}$
necessary for instability   becomes infinitely small.

The maximum fields allowing for  the axisymmetric S0-modes and the
nonaxisymmetric S1-modes in Fig.~\ref{f2} are almost equal. This property might be important for  the dynamo theory with respect to the Cowling theorem.

Equations (\ref{3.1}) - (\ref{3.3}) correspond  to  (\ref{1.14})  obtained with  the  local analysis. Quantitative differences are partly due to the difference in definitions of Rm between local and global calculations. The differences are actually quite small if  the transformation  rule $k \rightarrow \pi/H$ between local and global formulations is used.
\begin{figure}
   \begin{minipage}[c]{4.35 truecm}
      \includegraphics[width=4.3truecm]{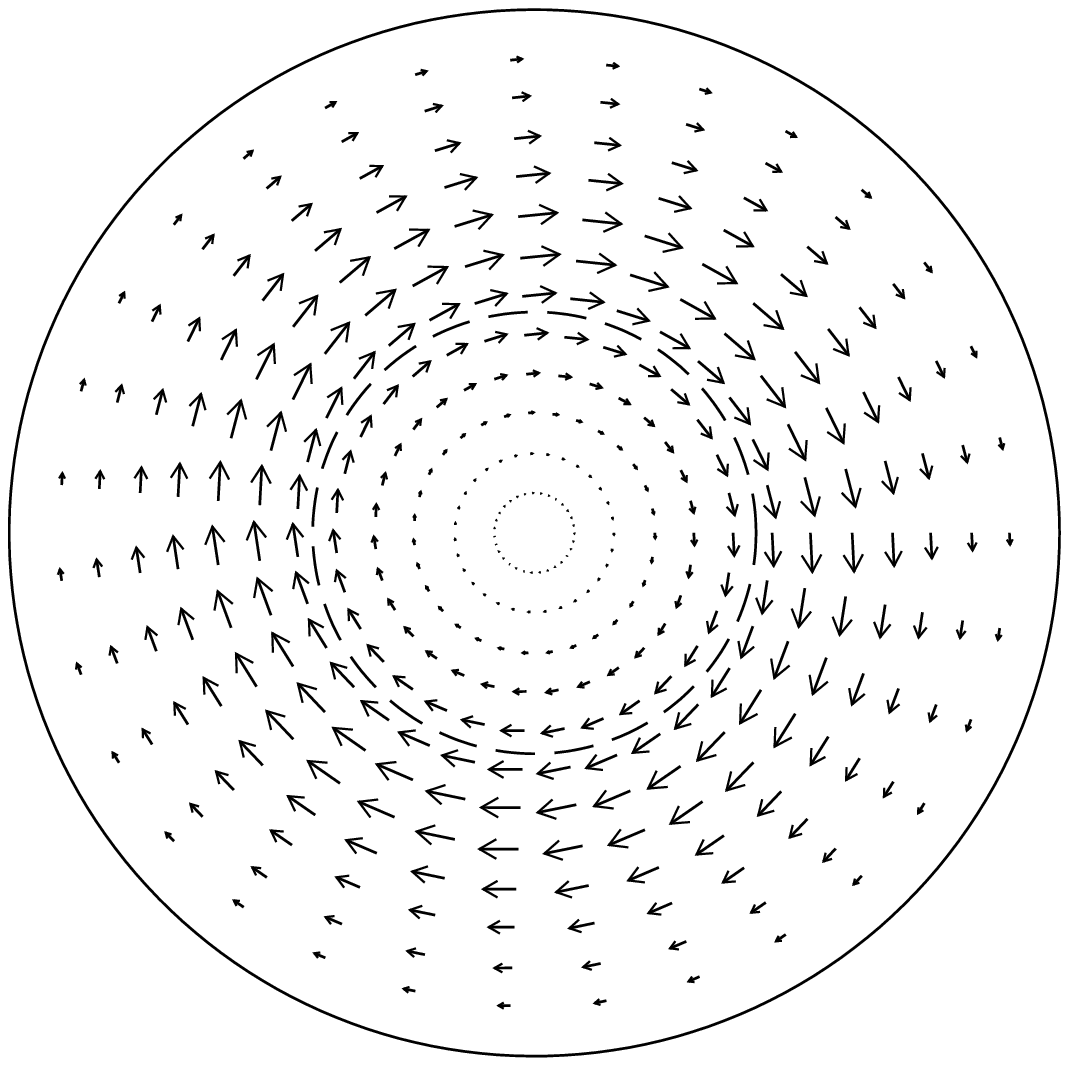}
   \end{minipage}
   \begin{minipage}[c]{4.35 truecm}
      \includegraphics[width=4.3truecm]{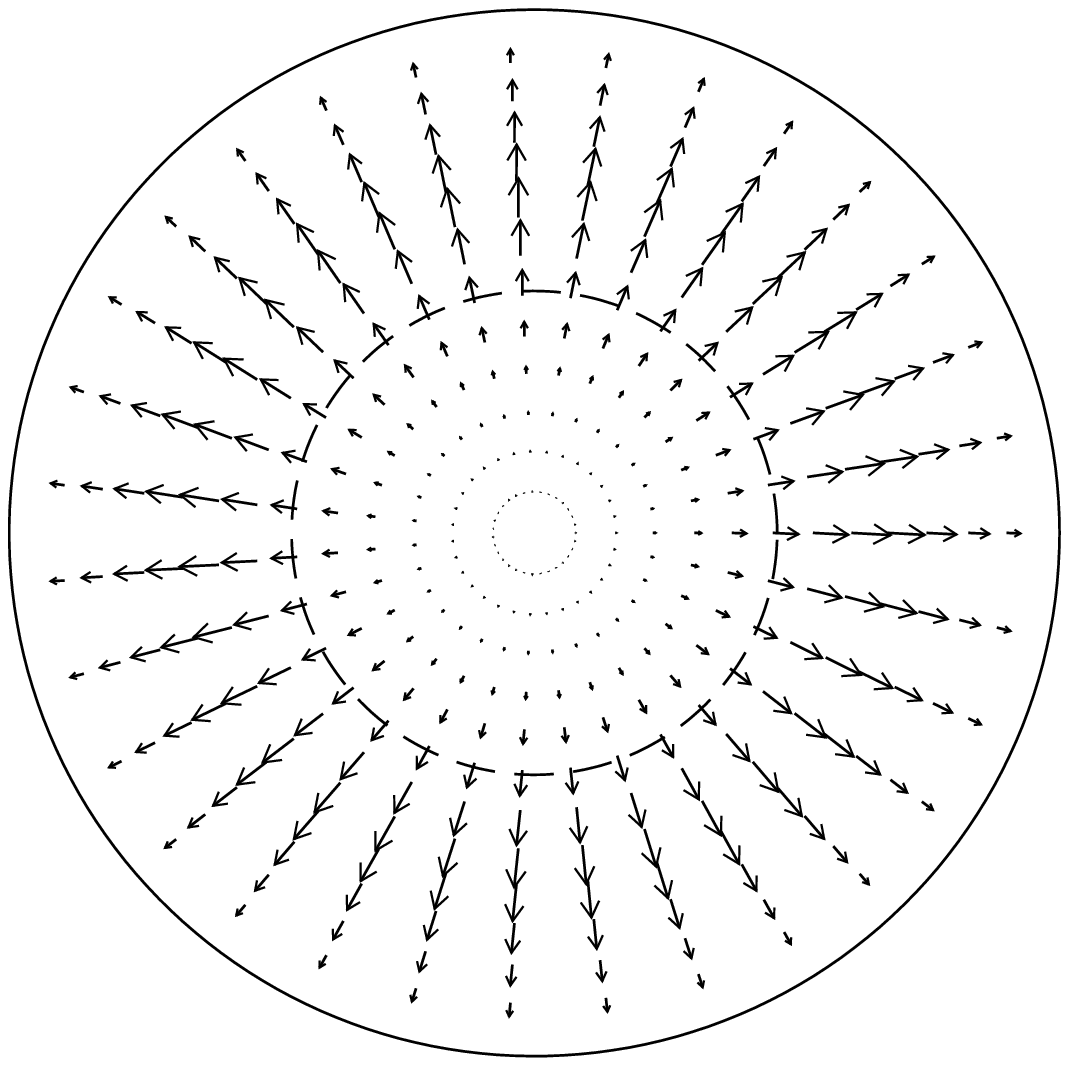}
   \end{minipage}
   \caption{Magnetic field vectors in the  disk midplane 
            for marginally stable S0-modes at ${\rm Rm}=100$  for the minimum (left) 
	    and maximum (right) Lundquist numbers, resp.
	    The broken circle shows the turnover radius $s_0$ of the rotation law.}
   \label{f3}
\end{figure}

Figure~\ref{f3} shows that the pitch angle for marginally stable disturbances is
indeed very small for the minimum field producing the instability. The angle
increases to about $\pi/2$ for the maximum field. This tendency for the pitch
angle can also explain the results for nonaxisymmetric disturbances. For pitch
angle close to $\pi/2$ the azimuthal structure is not significant. Accordingly,
the maximum fields for  the S1 and S0 modes   are roughly the same
(Fig.~\ref{f2}). Small pitch angle, however, would mean a tight winding with a
small radial scale for S1 modes whose instability is then precluded by
diffusion. This is  why the instability region of S1 modes in Fig.~\ref{f2} 
exhibits  such a sharp and almost vertical boundary on the weak-field side.

The orientation of the magnetic field in relation to the rotation axis plays 
an important role in the interplay of differential rotation  and the Hall effect.  Now the two cases of  parallel and antiparallel magnetic fields are considered solving the complete equations. We find that the MRI of the  Kepler flow  is  very differently modified by the Hall effect.
%%%%%%%%%%%%%%%%%%%%%%%%%%%%%%%%%%%%%%%%%%%%%%%%%%%%%%%%%%%%%%%%%%%%%%%
\subsection{Positive (parallel) fields}
\begin{figure}
   \centering
   \includegraphics[height=6cm,width=7cm]{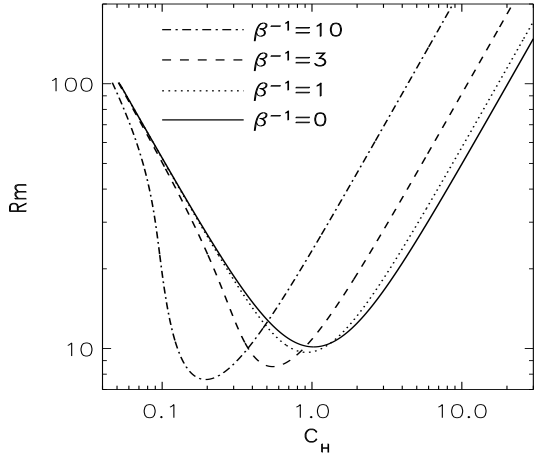}
   \caption{The same as in Fig. \ref{f2} but for the full equation
            system.  The Hall effect {\em increases} the
	    minimum Rm. The solid curve for $\beta \to \infty$
	    is identical to the lower curve in Fig. \ref{f2}.
	    For increasing Rm the lower limit for the magnetic field
	    becomes smaller and smaller.}
   \label{f6}
\end{figure}

%%%%%%%%%%%%%%%%%%%%%%%%%%%%%%%%%%%%%%%%%%%%%%%%%%%%%%%%%%%%%%%%%%%%%%%
 For positive $\beta$ the solution is located between the two realizations
given in  Figs.~\ref{f4} and \ref{f2}.  The minimum magnetic Reynolds number  
moves from the value 7 for MRI to about 10 for the shear-Hall instability. 
The Hall effect does thus  {\em not} support the instability of the cool  
Kepler flow; for all positive $\beta$ the minimum remains between 7 and 10 (Fig. \ref{f6}).

Figure~\ref{f6} also shows that  the  minimum  of the possible magnetic fields is fixed by the Hall effect. We find ${\rm Rm} \sim 1/ (\beta \cdot {\rm S})$ so that ${\rm S} \sim 1/(\beta \cdot {\rm Rm})$ which for large Rm is {\em smaller by orders of 
magnitude} than  ${\rm S}\sim 1$ taken from Fig. \ref{f5} for negative $\beta$.

Our main interest also concerns the results  for low electrical conductivity, i.e. for small Rm. The minimum Rm  moves in Fig. \ref{f6} from 7.1 to ~10, i.e. it increases  opposite to the desired trend.
If the results for parallel and antiparallel magnetic fields are compared, one of the differences   is the opposite trend for the minimum magnetic Reynolds number which  only for negative  fields (antiparallel to the rotation axis) is reduced by the Hall effect. It is also important that for large enough Rm  the  minimum magnetic fields for the instability   strongly differ for both the magnetic orientations.
%%%%%%%%%%%%%%%%%%%%%%%%%%%%%%%%%%%%%%%%%%%%%%%%%%%%%%%%%%%%%%%%%%%%%%%
\subsection{Negative (antiparallel) fields}
\begin{figure}
   \centering
   \includegraphics[height=6cm,width=7cm]{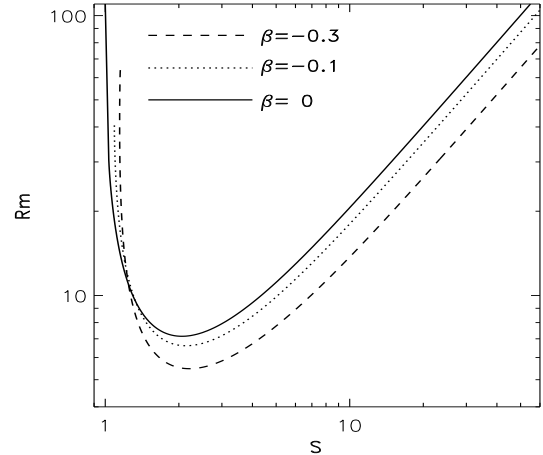}
   \caption{Stability diagram for MRI modified by the Hall effect
            for magnetic fields {\em antiparallel} to the angular
            velocity. Note that the Hall parameter  $\beta$  for real cool
            protostellar disks is of order  unity.
            The Hall effect shifts the instability interval to larger Lundquist numbers, i.e. to higher magnetic fields.}
   \label{f5}
\end{figure}

%%%%%%%%%%%%%%%%%%%%%%%%%%%%%%%%%%%%%%%%%%%%%%%%%%%%%%%%%%%%%%%%%%%%%%%
Figure~\ref{f5} shows the stability diagram for negative $B_0$ for which the 
shear-Hall instability does not exist. The Hall effect, however, increases both instability limits  of the field  amplitudes. The Hall effect  thus destabilizes for  strong fields close to $B_{\rm max}$ and stabilizes close to $B_{\rm min}$.

On the other hand, the absolute minimum of the  magnetic Reynolds number Rm is 
reduced. The necessary electrical conductivity of the gas is (slightly) 
reduced by the Hall effect. From numerical arguements, we can present only the 
results for a small Hall parameter $\beta$ which is still too small by  one order of magnitude  (see Eq. \ref{4.5} below). A massive reduction of the critical magnetic Reynolds number (i.e. the necessary electrical conductivity) is thus expected from the Hall effect for antiparallel magnetic fields. It seems indeed to be possible to use the Hall effect to realize the MRI also for the rather low electrical conductivity of cool disks.
%%%%%%%%%%%%%%%%%%%%%%%%%%%%%%%%%%%%%%%%%%%%%%%%%%%%%%%%%%%%%%%%%%%%%%%
\subsection{Energy relation and angular momentum transport}
%%%%%%%%%%%%%%%%%%%%%%%%%%%%%%%%%%%%%%%%%%%%%%%%%%%%%%%%%%%%%%%%%%%%%%%
Linear computations cannot provide energy values. It is possible,  however, to compare their global magnetic and kinetic energies. These ratios of the total (volume integrated) energies are given in Table~\ref{t1} for a sequence of values of the Hall parameter $\beta$.
The total energy is normally dominated by its magnetic part in agreement with other simulations of MRI (cf. Stone \& Norman 1994; Brandenburg et al. 1995; Sano \& Stone 2002). The energy ratio which we obtain is, however, decreasing and even drops below unity when the Hall parameter becomes 
more negative. A similar trend can also be found in the simulation results of
Sano \& Stone (2002). The angular momentum always flows outwards, and it is 
dominated by the Maxwell stress (Balbus \& Hawley 1991; Brandenburg et al.
1996). The relative contribution of the Reynolds stress also increases with 
decreasing $\beta$, in accordance with  the results of Sano \& Stone (2002,
their Table 4).
\begin{table}
\caption{Ratio of the global magnetic to kinetic energy at minimum
         magnetic Reynolds number for a sequence of the Hall
         parameter $\beta$.}
\begin{tabular}{c|llllll}
  \hline
  $\beta$ & $-0.3$ & $-0.1$ & 0 & 0.1 & 0.33 & 1 \\
  $E_{\rm mag}/E_{\rm kin}$ & 0.77 & 1.20 & 1.52 & 1.94 & 3.32 & 11.4 \\
  \hline
\end{tabular}
\label{t1}
\end{table}
%%%%%%%%%%%%%%%%%%%%%%%%%%%%%%%%%%%%%%%%%%%%%%%%%%%%%%%%%%%%%%%%%%%%%%%
\section{Discussion}\label{proto}
%%%%%%%%%%%%%%%%%%%%%%%%%%%%%%%%%%%%%%%%%%%%%%%%%%%%%%%%%%%%%%%%%%%%%%%
Ionization of protostellar disks material
should be extremely low (Gammie 1996). The electrical conductivity is then controlled by electron-neutral collisions and the magnetic diffusivity 
\begin{equation}
  \eta = 234\ {n\over n_{\rm e}}\ T^{1/2}\ {\rm cm}^2\ {\rm s}^{-1} 
\label{4.1}
\end{equation}
is inversely proportionate to the ionization ratio (cf. Balbus \& Terquem 2001)
where $n_{\rm e}$ and $n$ are the number densities of electrons and neutrals. Then the magnetic Reynolds number (\ref{2.5}) can be estimated with Eq.~(\ref{4.1}) as
\begin{equation}
  {\rm Rm}\ =\ 2\cdot 10^{15} \left({n_{\rm e}\over n}\right)
  \left({T\over 100\ {\rm K}}\right)^{-1/2}
  \left({\tau_{\rm rot}\over 0.1{\rm yr}}\right)^{-1}
  \left({H\over 0.1{\rm AU}}\right)^2 ,
\label{4.2}
\end{equation}
where $\tau_{\rm rot}$ is the rotation period at the distance $s_0 = 5 H$.

The ionization fraction, $n_{\rm e}/n$, is very uncertain. We estimate first its value required for instability.
Without the Hall effect our results suggest ${\rm Rm} \simeq 10$ for instability. Assuming numerical values in  Eq.~(\ref{4.2}) as representative, one finds
\begin{equation}
  {n_{\rm e}\over n } \simeq 10^{-14}
  \label{4.3}
\end{equation}
as the critical value. Even such a low ionization is problematic for
protostellar disks (Stepinski 1992; Gammie 1996). Collisional ionization is inefficient for $T < 10^3$~K. The cosmic ray, however, can provide ionization fractions of
$n_{\rm e}/n \simeq 10^{-12}$
if the column density $\Sigma$  does not exceed $10^2$~g~cm$^{-2}$ (Umebayashi \& Nakano 1981; Gammie 1996). With this value, Eq.~(\ref{4.2}) for thick disks ($H\simeq 0.1$ AU)
provides Rm~$\simeq 10^3$  but this value reduces to 10  for thinner disks ($H\simeq 0.01$ AU).

\begin{figure}
   \centering
   \includegraphics[height=7.2cm,width=8.5cm]{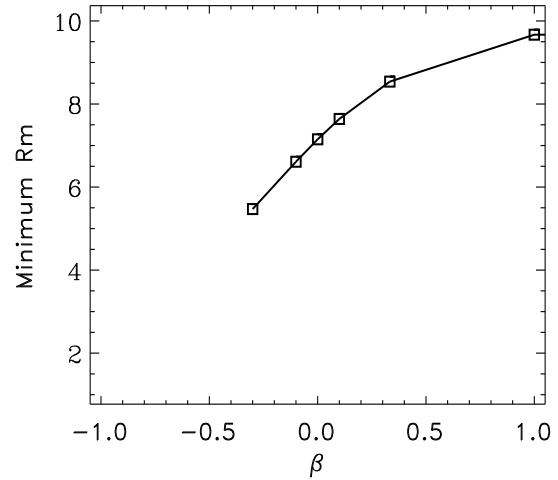}
   \caption{The computed minimum magnetic Reynolds number as a function of the Hall parameter. $\beta=0$  denotes MRI.  The  results for  $\beta<-0.3$  are not yet known.}
   \label{f7}
\end{figure}

The Hall parameter (\ref{1.3}) can be estimated as
\begin{equation}
  C_{\rm H} \simeq 20
  \left({T\over 100\ {\rm K}}\right)^{-1/2}
  \left({n\over 10^{14}{\rm cm}^{-3}}\right)^{-1} {B\over 1 G},
  \label{4.4}
\end{equation}
and $\beta$ reads
$\beta$ of Eq.~(\ref{1.20}), reads
\begin{equation}
  |\beta|\ \simeq 2\cdot 10^{-12} {n\over n_{\rm e}}
  \left( n\over 10^{14}{\rm cm}^{-3}\right)^{-1/2}
  \left( H\over 0.1{\rm AU}\right)^{-1} .
  \label{4.5}
\end{equation}
With a particle density of $n\simeq 10^{14}$~cm$^{-3}$, this relation  leads to the value of 
\begin{equation}
|\beta|\simeq 2,
\end{equation}
so  that the Hall effect should indeed be very important. With such  large 
values one can take from Fig.~\ref{f6}  that the pure shear-Hall instability 
dominates for positive fields. In  Fig.~\ref{f7} the numerical results are 
summarized for the dependence  of the critical magnetic Reynolds number on the Hall parameter $\beta$.  For negative $\beta$ of order unity it is obvious that the critical magnetic Reynolds number is reduced by at least one order  of magnitude by the Hall effect.

The  boundaries  of the instability domain also strongly depend on the field 
orientation. Only for antiparallel fields does an instability exist for $\rm Rm <7$. If the magnetic Reynolds number exceeds O(10) then both magnetic orientations lead to instability but for rather different magnetic amplitudes. It is
\begin{equation}
   0.1\ {\rm G} < B_0 < 10\ {\rm G}
   \label{4.6}
\end{equation}
for antiparallel fields and
\begin{equation}
   0.001\ {\rm G} < B_0 < 1\ {\rm G}
   \label{4.7}
\end{equation}
for parallel fields, both taken for $\rm Rm =100$. Note that the magnetic 
fields of meteorites vary between 1 G and 10 G, close to the upper limits of 
the above equations. This coincidence suggests that the instability may drive a
dynamo which saturates when the field is amplified to the upper boundary of the
instability interval. Also, the minimum magnetic fields allowing the 
instability  prove to be rather strong. If the external magnetic fields never 
exceed 0.1 G  in amplitude then only the Hall effect of  parallel fields would 
lead to the instability necessary to remove the angular momentum in the Kepler disk but then the ionization must be high enough. If it is not, then  the Hall effect for antiparallel fields is needed but  the magnetic field {\em must} exceed 0.1 G in this case. It this clear from such considerations 
%that the star formation rate in dense and cold globules of molecular clouds 
that in dense and cold globules of molecular clouds there are severe limitations to the necessary transport of  angular momentum by magnetic instabilities.

%%%%%%%%%%%%%%%%%%%%%%%%%%%%%%%%%%%%%%%%%%%%%%%%%%%%%%%%%%%%%%%%%%%%%%%
\begin{acknowledgements}
LLK is grateful to the Astrophysical Institute Potsdam  for its hospitality and the visitor support.
\end{acknowledgements}
%%%%%%%%%%%%%%%%%%%%%%%%%%%%%%%%%%%%%%%%%%%%%%%%%%%%%%%%%%%%%%%%%%%%%%%
\bibliographystyle{aa}
%%%%%%%%%%%%%%%%%%%%%%%%%%%%%%%%%%%%%%%%%%%%%%%%%%%%%%%%%%%%%%%%%%%%%%%

%%%%%%%%%%%%%%%%%%%%%%%%%%%%%%%%%%%%%%%%%%%%%%%%%%%%%%%%%%%%%%%%%%%%%%%
\end{document}